\documentclass[%
 reprint,
showpacs,preprintnumbers,
nofootinbib,
 amsmath,amssymb,
 aps,
 showkeys,
 showpacs
]{revtex4-1}
\usepackage{graphics}
\usepackage{bbold}
\usepackage{color}

\usepackage{cancel}
\usepackage[normalem]{ulem}

\usepackage{array}
\usepackage{booktabs}
\usepackage{amsmath}
\usepackage{graphicx}
\usepackage{subfigure}

\usepackage{amsmath}
\usepackage{slashed}

\usepackage{multirow}

\usepackage{float}

\usepackage{lineno,hyperref}
\def\be		{\begin{eqnarray}}
\def\en		{\end{eqnarray}}
\def\nen	{\nonumber\end{eqnarray}}

\def\lt		{\left(}
\def\rt		{\right)}

\def\jp		{\ensuremath{J\!/\psi}}
\def\psii		{\ensuremath{\psi(2S)}}

\def\ee		{\ensuremath{e^+e^-}}
\def\BB		{\ensuremath{B\bar{B}}}

\def\ampl   {\ensuremath{\mathcal A}}
\def\De     {\ensuremath{D_e}}

\def\Fe     {\ensuremath{F_e}}

\definecolor{amber}{rgb}{1.0, 0.55, 0.0}
\definecolor{redmod}{rgb}{0.89, 0.0, 0.13}
\definecolor{blumod}{rgb}{0.01, 0.28, 1.0}

\def \br {{\rm{BR}}}

\begin{document}

\title{
The electromagnetic amplitudes in the $J/\psi$ and $\psi(2S)$ decays into spin-1/2 baryon-antibaryon pairs
}
\author{Alessio Mangoni}
\affiliation{%
INFN Sezione di Perugia, I-06100, Perugia, Italy}%%
\begin{abstract}
After investigating the decays of a charmonium $\psi = J/\psi, \psi(2S)$ into a spin-1/2 baryon-antibaryon $\BB$ pair, with the determination of the only parameter that gives the EM amplitude for neutral final states, in this work we focus our attention on the decays into charged baryons, whose EM amplitudes can be expressed in terms of a further parameter.\\
By considering the BESIII data on the $e^+e^- \to p \overline p$ cross section we obtain a full parametrization for the EM amplitudes and make predictions on the cross section of the decays $e^+ e^- \to \Sigma^+ \overline \Sigma{}^-$, $e^+ e^- \to \Sigma^- \overline \Sigma{}^+$ and $e^+ e^- \to \Xi^- \overline \Xi{}^+$ at the $J/\psi$ and $\psi(2S)$ masses.
\end{abstract}

\maketitle

\section{Introduction}
\label{sec:intro}
\begin{figure}[H]
\centering
\includegraphics[width=0.65\columnwidth]{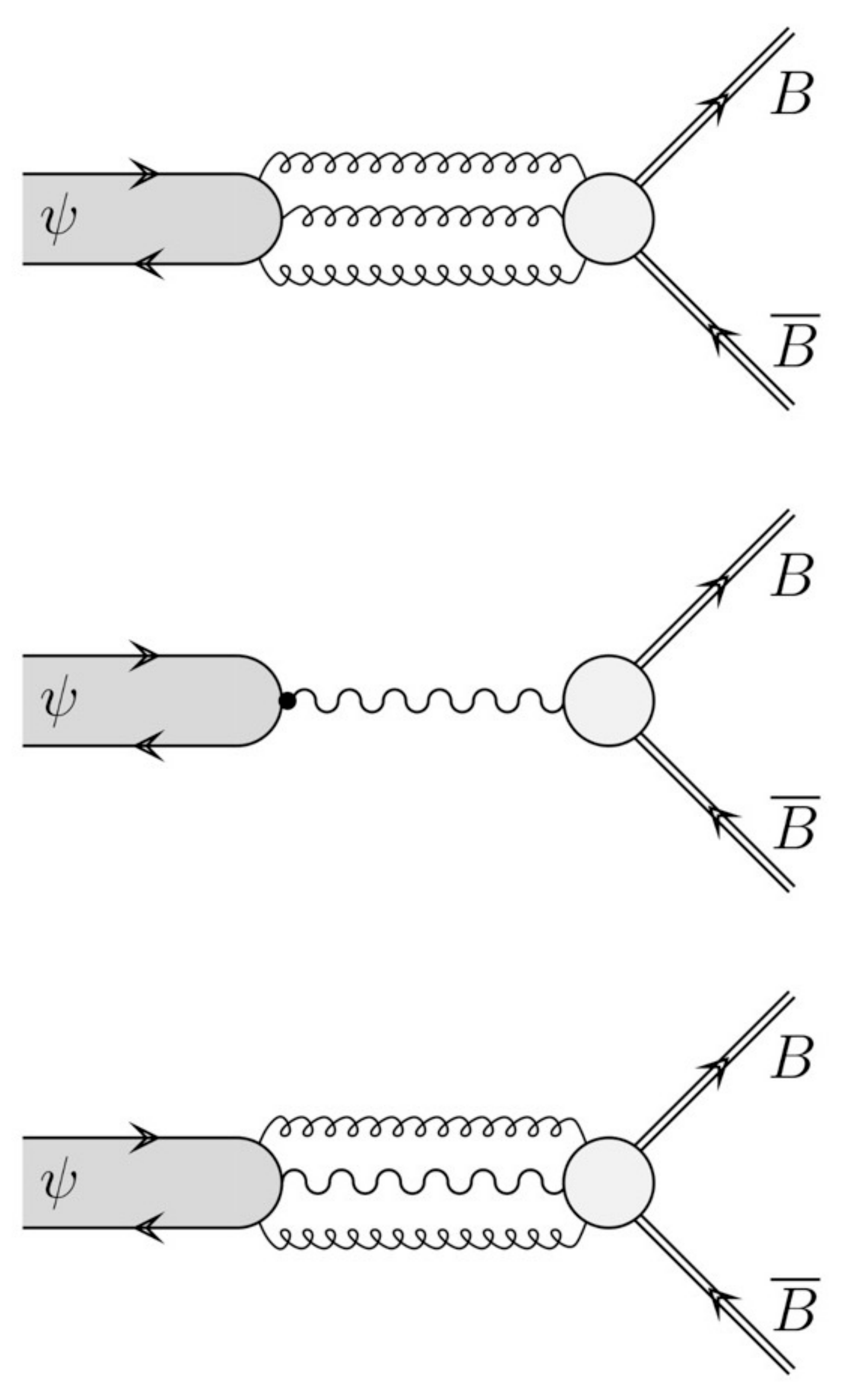}
\caption{\label{fig.ampls} Main contributions for the decay $\psi \to \BB$.}
\end{figure}

The decays $\psi \to \BB$, being $\psi$ a vector charmonia, $\psi=J/\psi, \psii$, where $B$ is a spin-1/2 baryon of the SU(3) octet, represented by the matrix
\be
B=\begin{pmatrix}
\Lambda/\sqrt{6}+\Sigma^0/\sqrt{2} & \Sigma^+ & p\\
\Sigma^- & \Lambda/\sqrt{6}-\Sigma^0/\sqrt{2} & n\\
\Xi^- & \Xi^0 & -2 \Lambda/\sqrt{6}\end{pmatrix} \,,
\nen
have been studied from decades~\cite{Kopke:1988cs,Claudson:1981fj}.\\
The generic decay $\psi \to \BB$ can be written as the sum of three main contributions: strong, electromagnetic (EM) and mixed strong-EM, as shown in Figure~\ref{fig.ampls}, where, in case of perturbative QCD (pQCD), the latter is proportional to the former~\cite{Korner:1986vi}. \\
The three related amplitudes were studied and recently separated, for the first time, in Refs.~\cite{Ferroli:2019nex,Ferroli:2020mra}, for the $J/\psi$ and $\psi(2S)$, respectively.\\
The EM amplitudes for the neutral final states in the decays $\psi \to \BB$ can be parametrized~\cite{Ferroli:2020xnv} in terms of only one parameter, called $D_e$, while those of the charged final states can be expressed in terms of the parameter $D_e$ and the new parameter $F_e$. \\
The first parameter, $D_e$ have been deeply investigated in Ref.~\cite{Ferroli:2020xnv}, with the determination of its modulus $|D_e|$.\\
In this work we focus our attention on the $F_e$ parameter, related to charged baryon final states, i.e., we consider the cases where $\BB \in \{ p \overline p, \Sigma^+ \overline \Sigma{}^-, \Sigma^- \overline \Sigma{}^+, \Xi^- \overline \Xi{}^+ \}$. 
%%%
%%%

\section{Cross sections and branching ratios}
The electromagnetic amplitudes for the decays $\psi \to \BB$ can be parametrized in terms of few parameters for the whole SU(3) octet.
Following the procedure discussed in Ref.~\cite{Ferroli:2020xnv} the EM branching ratio of the generic decay $\psi \to \gamma^* \to \BB$ can be written as~\cite{Ferroli:2020xnv}
\be
%\label{eq.Brgen1}
\br_{\BB}^\gamma = {\beta_{M_B}(M_\psi^2) \over 16 \pi M_{\psi} \Gamma_{\psi}}  \,|g^\psi_{\gamma}\ampl^\gamma_{\BB}(M_\psi^2)|^2 \,,
\nen
where
\be
%\label{eq:velo}
\beta_{M_B}(q^2)=\sqrt{1- {4M_B^2 \over q^2}}
\nen
is the velocity of the outgoing baryon in the $\BB$ center-of-mass frame and $g^\psi_{\gamma}$ is the coupling constant between the $\psi$ meson and the virtual photon $\gamma^*$.\\
The amplitudes parametrizations are reported in Table~\ref{tab:A.par.gamma}.
\begin{table}[h]
%\vspace{-2mm}
\begin{center}
\caption{Parametrizations of the amplitudes of the EM decay $\psi\to\gamma^*\to\BB$ as function of the couplings $D_e$ and $F_e$~\cite{Ferroli:2020xnv}.}
\label{tab:A.par.gamma} 
\begin{tabular}{lcl} 
\hline\noalign{\smallskip}
\BB &\hspace{10mm} & $g^\psi_\gamma\ampl^\gamma_{\BB}(M_\psi^2)$ \\
\noalign{\smallskip}\hline\noalign{\smallskip}
$\Sigma^0 \overline \Sigma{}^0$ && $\De$ \\
$\Lambda \overline \Lambda$ && $-\De$ \\
$\Lambda \overline \Sigma{}^0+ \rm{c.c.}$ && $\sqrt{3}\,\De$ \\
$p \overline p$ && $\De + \Fe$ \\
$n \overline n$ && $-2\,\De$ \\
$\Sigma^+ \overline \Sigma{}^-$ && $\De + \Fe$ \\
$\Sigma^- \overline \Sigma{}^+$ && $\De - \Fe$ \\
$\Xi^- \overline \Xi{}^+$ && $\De - \Fe$ \\
$\Xi^0 \overline \Xi{}^0$ && $-2\,\De$ \\
\noalign{\smallskip}\hline
\end{tabular}
\end{center}
\end{table} \\
The moduli of the parameter $D_e$, for both $\jp$ and $\psii$ mesons, have been determined recently as~\cite{Ferroli:2020xnv}
\begin{gather}
\label{eq.De-val-ref}
\begin{aligned}
|D_e|_{\jp} = (3.93 \pm 0.17) \times 10^{-4} \ {\rm GeV}\,,  \\
|D_e|_{\psii} = (1.25 \pm 0.07) \times 10^{-4} \ {\rm GeV}\,. 
\end{aligned}
\end{gather}
The Born cross section $e^+e^- \to \BB$ at the $\psi$ mass, $q^2 = M_\psi^2$, can be written in the form~\cite{Ferroli:2020xnv}
\be
\label{eq.cs2}
\sigma_{\BB} (M_\psi^2) &=& {4 \pi \alpha^2 \over 3M_\psi^2 \br_{\mu^+ \mu^-}^\gamma} \, \br^\gamma_{ \BB} \nonumber \\
&=& {\left|g^\psi_{\gamma}\right|^2 \alpha^2 \beta_{M_B}(M_\psi^2) \over 12M_\psi^3 \Gamma_{\psi} \br_{\mu^+ \mu^-}^\gamma} \,|\ampl^\gamma_{\BB}(M_\psi^2)|^2 \,,
\en
where
\be
\br_{\mu^+ \mu^-}^\gamma = {|g^\psi_{\gamma}|^2 \over 16 \pi M_\psi \Gamma_\psi}
\nen
is the BR of the decay $\psi \to \mu^+ \mu^-$.\\
Concerning one of the four final states with charged baryon, the $p \overline p$ one, several recent data are available on the cross section $e^+ e^- \to p \overline p$, from the BESIII collaboration~\cite{Ablikim:2019njl}.\\
In this case the cross section of Eq.~\eqref{eq.cs2}, using the parametrization reported in Table~\ref{tab:A.par.gamma}, becomes
\be
\label{eq.cspp}
\sigma_{p \overline p} (M_\psi^2) = {\alpha^2 \beta_{M_p}(M_\psi^2) \over 12M_\psi^3 \Gamma_{\psi} \br_{\mu^+ \mu^-}^\gamma} \,|D_e+F_e|^2 \,,
\en
where, from PDG~\cite{Zyla:2020zbs},
\be
\br_{\mu^+ \mu^-}^\gamma = (5.961 \pm 0.033) \times 10^{-2} \,.
\nen
%To facilitate the calculations we define also a scaled cross section
%\be
%\label{eq.scaled-not-scaled}
%\tilde \sigma_{p \overline p} (M_\psi^2) = {\sigma_{p \overline p} (M_\psi^2) \over \beta_{M_p}(M_\psi^2)} \,,
%\en
%obtained from the ratio of the cross section and the velocity of the outgoing proton.
%
%
%
\section{Results}
\label{sec:scaled-ee}
\begin{figure}[H]
\includegraphics[width=0.99\columnwidth]{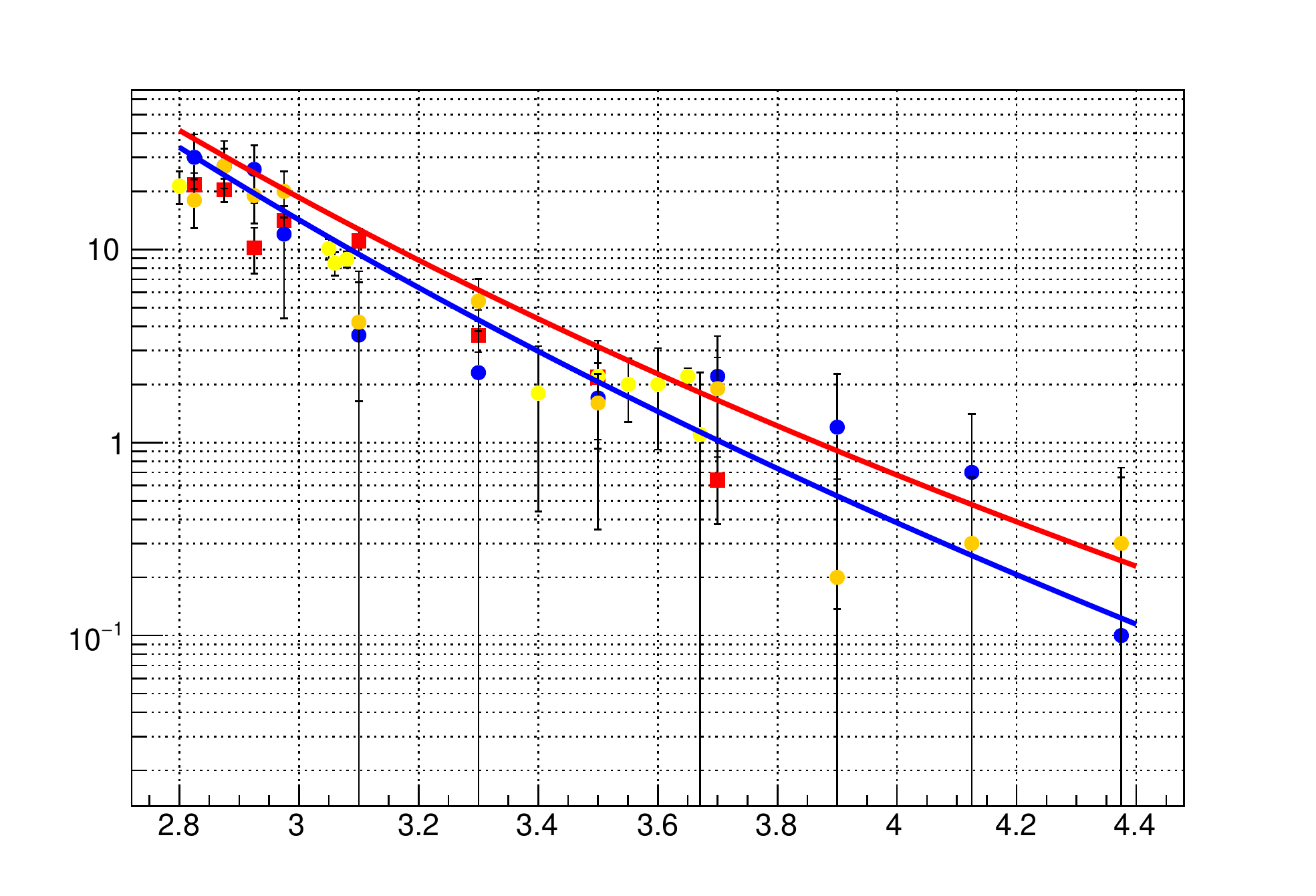} 
\put(-220,157){$\sigma_{p \overline p} \ \rm [pb]$}
\put(-67,-3){$\sqrt{q^2} \ \rm [GeV]$}
\caption{\label{fig.data-fit-pp} Plot of the $e^+e^- \to p \overline p$ cross section in logarithmic $y$-scale. The points are the data on the cross section from BESIII~\cite{Ablikim:2019njl} (red), \cite{Ablikim:2015vga} (yellow) and BABAR~\cite{Aubert:2005cb} (blue), \cite{Lees:2013ebn} (orange), while the red and blue bands represent the fitting functions of Eq.~\eqref{eq.fit-p1} and~\eqref{eq.fit-p2}, respectively.}
\end{figure}

We use the newest BESIII~\cite{Ablikim:2019njl} data on the cross section of $\ee\to p \overline p$ to determine the modulus $|D_e+F_e|$ of Eq.~\eqref{eq.cspp}. We consider the two fit functions for the cross section from Ref.~\cite{Ablikim:2019njl}. The first one is based on a QCD-inspired parametrization of the form factors~\cite{Bianconi:2015vva,Shirkov:1997wi}
\be
\label{eq.fit-p1}
\sigma_{\rm fit,1}(q^2) = {A_1 \left( 1+2M_p^2/q^2 \right) \over (q^{2})^{5} \lt\pi^2 + \ln^2( q^2 / \Lambda_{\rm QCD}^2 ) \rt^2 } \,,
\en
that includes also the logarithmic corrections, while the second is suggested by Ref.~\cite{TomasiGustafsson:2001za}
\be
\label{eq.fit-p2}
\sigma_{\rm fit,2}(q^2) = {A_2 \left( 1+2M_p^2/q^2 \right) \over q^{2} (1-q^2/q_0^2)^4 (1+q^2/m_a^2)^2 } \,,
\en
where $M_p$ is the proton mass, $A_1=72 \ \rm GeV^4$ and $A_2=7.7$, while $\Lambda_{\rm QCD} = 0.52 \ \rm GeV$, $q_0^2 = 0.71 \ \rm GeV^2$ and $m_a^2 = 14.8 \ \rm GeV^2$~\cite{Ablikim:2019njl}.
\\
%The fit has been performed on the cross section data points, obtained by the BESIII and the BABAR experiments, considering only data at $ q^2 \ge (3\,{\rm GeV})^2$, to avoid the threshold energy region.\\
%The obtained values for the parameters $A_1$ and $A_2$ from the minimization procedure are
%\begin{gather}
%\label{eq.A.obt1}
%\begin{aligned}
%A_1=(2.82 \pm 0.14)\times 10^{8} \ \rm GeV^{10} \, pb\,.\\
%A_2=(4.14 \pm 0.23)\times 10^{6} \ \rm GeV^{2} \, pb\,.
%\end{aligned}
%\end{gather}
In Figure~\ref{fig.data-fit-pp} are shown the cross section data from various experiments, together with the fitting function of Eq.~\eqref{eq.fit-p1} and~\eqref{eq.fit-p2}.\\
In order to obtain the values of $|D_e+F_e|$, using Eq.~\eqref{eq.cspp}, we combine\footnote{the final value is given by the mean of the two values, while its error is calculated as their standard deviation.} the results for the cross sections of the two fitting functions of Eq.~\eqref{eq.fit-p1} and~\eqref{eq.fit-p2}. The values are
\begin{gather}
\label{eq.DepFe}
\begin{aligned}
|D_e+F_e|_{\jp} = (11.6 \pm 1.2) \times 10^{-4} \ \rm GeV\,, \\
|D_e+F_e|_{\psii} = (3.31 \pm 0.55) \times 10^{-4} \ \rm GeV\,,
\end{aligned}
\end{gather}
%%%
%%%
%\begin{table} [H]
%\vspace{-2mm}
%\centering
%\caption{Cross section computed through Eq.~\eqref{eq.scaled-not-scaled}, using the Eq.~\eqref{eq.fit-p} for the scaled cross section with the value of $A$ given in Eq.~\eqref{eq.A.obt}.}
%\label{tab:br-in-ou-pp} 
%\begin{tabular}{lr} 
%\hline\noalign{\smallskip}
%Energy & Cross section $\sigma_{p \overline p}$ \\
%$q^2=M_{J/\psi}^2$ & $(9.5 \pm 1.1) \ \rm pb$ \\
%$q^2=M_{\psi(2S)}^2$ & $(1.46 \pm 0.23) \ \rm pb$ \\
%\end{tabular}
%\end{table}
%
\begin{figure}[h!]
\includegraphics[width=0.95\columnwidth]{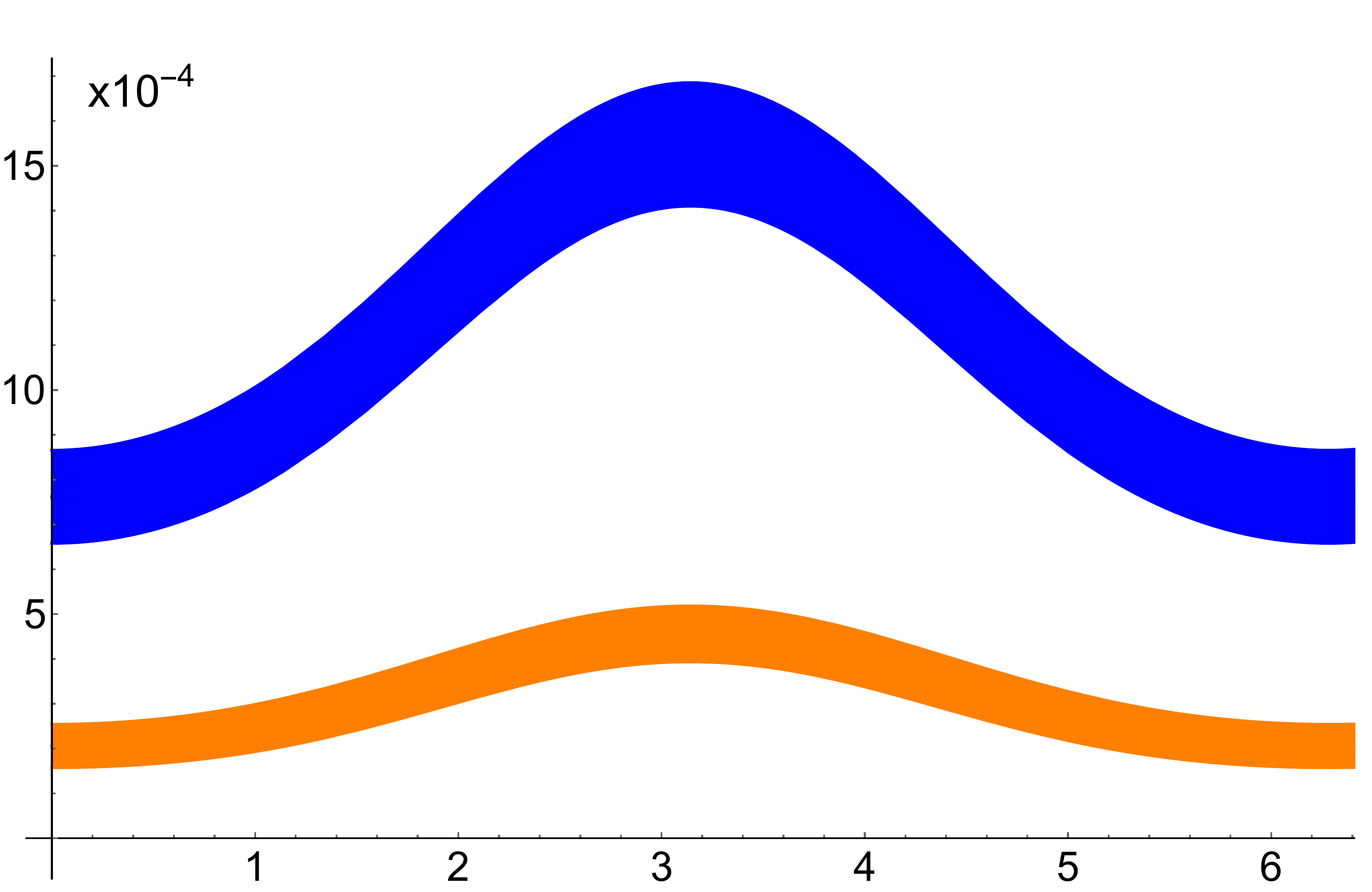}
\put(-234,150){$|F_e| \ \rm [GeV]$}
\put(-10,-1){$\rho$}
\caption{\label{fig:Fe-fase} Modulus of the $F_e$ parameter for the $J/\psi$ meson, blue band, and for the $\psi(2S)$, orange band, as a function of the relative phase, $\rho$, between $D_e$ and $F_e$.}
\end{figure}
\begin{figure}[h!]
\includegraphics[width=0.95\columnwidth]{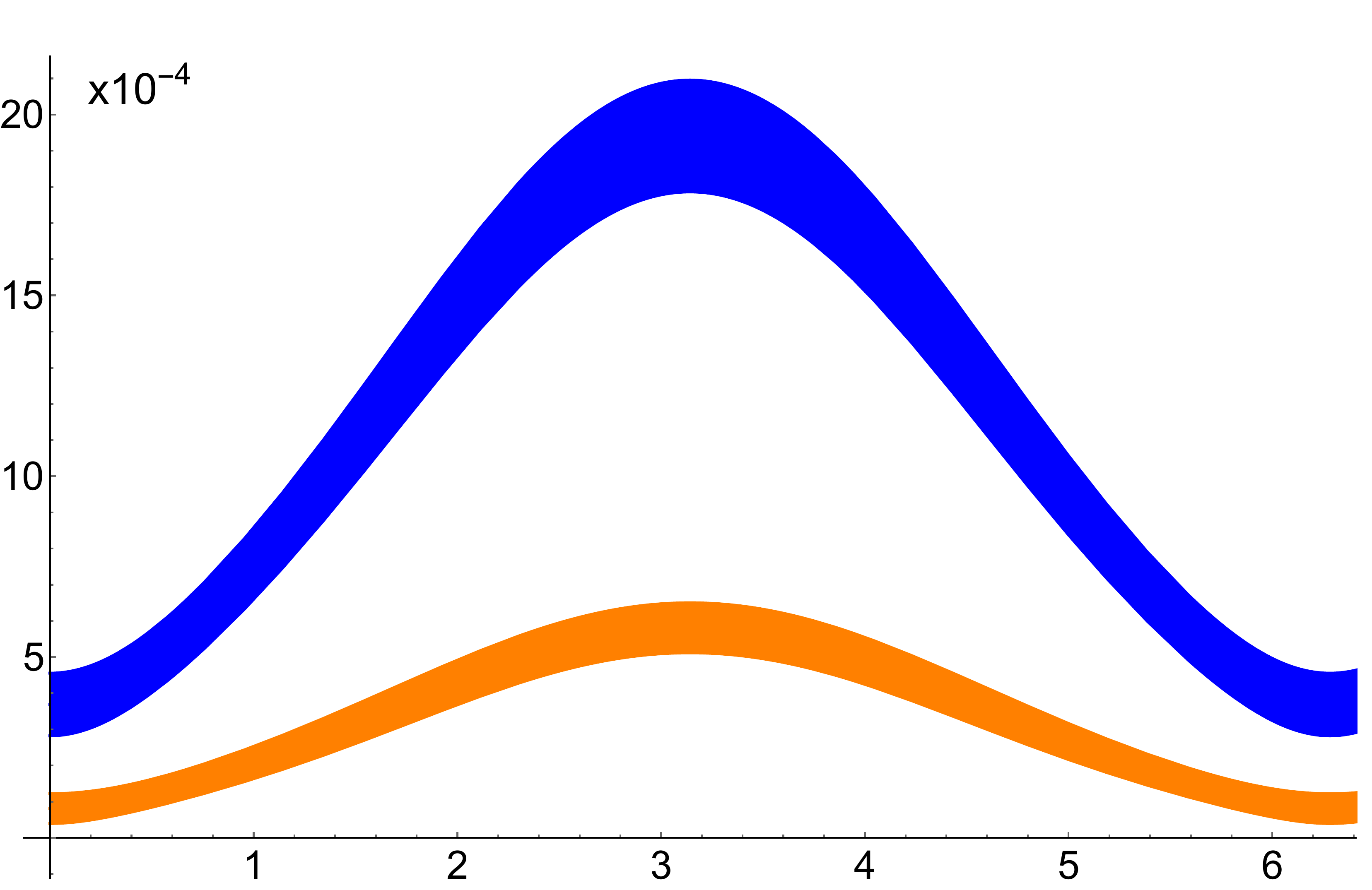}
\put(-234,150){$|D_e-F_e| \ \rm [GeV]$}
\put(-10,-1){$\rho$}
\caption{\label{fig:DemFe-fase} Modulus of $(D_e-F_e)$ for the $J/\psi$ meson, blue band, and for the $\psi(2S)$, orange band, as a function of the relative phase, $\rho$, between $D_e$ and $F_e$.}
\end{figure}
We can write the modulus of the parameter $F_e$ as a function of the obtained quantity $|D_e+F_e|$ and the relative phase $\rho$ between $D_e$ and $F_e$ as follows
\be
|F_e|_\psi = \sqrt{|D_e+F_e|_\psi^2\! -|\De|_\psi^2 \sin^2(\rho_\psi)} - |\De|_\psi \cos (\rho_\psi) \,.
\nen
A plot of $|F_e|$ as a function of the relative phase $\rho$ is shown in Figure~\ref{fig:Fe-fase}.\\
Using the obtained values of Eq.~\eqref{eq.DepFe}, and the values for the moduli of the $D_e$ parameter of Eq.~\eqref{eq.De-val-ref}, we can determine the following range of values for $|F_e|$
\begin{gather}
\label{eq.DepFe-val}
\begin{aligned}
6.58 \times 10^{-4} \ {\rm GeV} \leq |F_e|_{\jp} \leq 16.85 \times 10^{-4} \ {\rm GeV}\,, \nonumber \\
1.58 \times 10^{-4} \ {\rm GeV} \leq |F_e|_{\psii} \leq 5.18 \times 10^{-4} \ {\rm GeV}\,, \nonumber \\
\end{aligned}
\end{gather}
The quantity $|D_e-F_e|$ as a function of $|D_e+F_e|$, $|D_e|$ and $\rho$ is given by
\be
\label{eq.demfe}
|D_e-F_e| &=& \bigg(|D_e+F_e|^2 + 4 |D_e|^2 \cos^2 \rho - 4 |D_e| \cos \rho \nonumber \\
& & \times \sqrt{|D_e+F_e|^2-|D_e|^2 \sin^2 \rho} \bigg)^{1/2} \,. 
\en
and its plot is shown in Figure~\ref{fig:DemFe-fase}. Starting from this equation we can determine the range of values for $|D_e-F_e|$, i.e.,
\begin{gather}
\label{eq.DepFe-val}
\begin{aligned}
2.82 \times 10^{-4} \ {\rm GeV} \leq |D_e-F_e|_{\jp} \leq 20.95 \times 10^{-4} \ {\rm GeV}\,, \nonumber \\
0.40 \times 10^{-4} \ {\rm GeV} \leq |D_e-F_e|_{\psii} \leq 6.50 \times 10^{-4} \ {\rm GeV}\,, \nonumber \\
\end{aligned}
\end{gather}

\subsection{Real parameters}
Under the hypothesis that the two parameters $D_e$ and $F_e$ are relatively real, as suggested in Ref.~\cite{Ferroli:2019nex}, i.e., with $\rho=0$, we can determine the values of $|F_e|$ as
%%%
%\begin{table} [H]
%\vspace{-2mm}
%\centering
%\caption{Comparison between $D_e$ and $F_e$ related quantities, assuming that $D_e$ and $F_e$ are relatively real.}
%\label{tab:comp-de-fe-real-j} 
%\begin{tabular}{lrr} 
%\hline\noalign{\smallskip}
%Quantity & This work [$\rm GeV$] & Other work [$\rm GeV$] \\
%$10^4 \times |D_e|_{J/\psi}$ & - & $3.93 \pm 0.17$~\cite{Ferroli:2020xnv} \\ 
%$10^4 \times |D_e|_{\psi(2S)}$ & - & $1.25 \pm 0.07$~\cite{Ferroli:2020xnv} \\
%$10^4 \times |F_e|_{J/\psi}$ & $7.6 \pm 1.2$ & $7.91 \pm 0.62$~\cite{Ferroli:2019nex} \\ % discr 1.34
%$10^4 \times |F_e|_{\psi(2S)}$ & $2.06 \pm 0.56$ & $1.65 \pm 0.17$~\cite{Ferroli:2020mra} \\ % discr 1.67
%$10^4 \times |D_e+F_e|_{J/\psi}$ & $11.6 \pm 1.2$ & $12.43 \pm 0.65$~\cite{Ferroli:2019nex} \\ % discr 1.98
%$10^4 \times |D_e+F_e|_{\psi(2S)}$ & $3.31 \pm 0.55$ & $2.90 \pm 0.18$~\cite{Ferroli:2020mra} \\ % discr 1.64
%$10^4 \times |D_e-F_e|_{J/\psi}$ & $3.7 \pm 1.3$ & $3.39 \pm 0.65$~\cite{Ferroli:2019nex} \\ % discr 0.65
%$10^4 \times |D_e-F_e|_{\psi(2S)}$ & $0.81 \pm 0.57$ & $0.42 \pm 0.18$~\cite{Ferroli:2020mra} \\ % discr 1.56
%\end{tabular}
%\end{table}
\begin{table} [H]
\vspace{-2mm}
\centering
\caption{Comparison between $D_e$ and $F_e$ related quantities, assuming that $D_e$ and $F_e$ are relatively real. In the last column are indicated the discrepancies, in standard deviations (s.d.), between the values of second and third columns.}
\label{tab:comp-de-fe-real-j} 
\begin{tabular}{lrrr} 
\hline\noalign{\smallskip}
Quantity ($\times 10^{4}$) & This work [$\rm GeV$] & Other work [$\rm GeV$] & s.d. \\
\noalign{\smallskip}\hline\noalign{\smallskip}
$|D_e|_{J/\psi}$ & - & $3.93 \pm 0.17$~\cite{Ferroli:2020xnv} & - \\ 
$|D_e|_{\psi(2S)}$ & - & $1.25 \pm 0.07$~\cite{Ferroli:2020xnv} & - \\
$|F_e|_{J/\psi}$ & $7.6 \pm 1.2$ & $7.91 \pm 0.62$~\cite{Ferroli:2019nex} & $\sim 0.2$ \\ % 
$|F_e|_{\psi(2S)}$ & $2.06 \pm 0.56$ & $1.65 \pm 0.17$~\cite{Ferroli:2020mra} & $\sim 0.7$ \\ % 
$|D_e+F_e|_{J/\psi}$ & $11.6 \pm 1.2$ & $12.43 \pm 0.65$~\cite{Ferroli:2019nex} & $\sim 0.6$ \\ % 
%$10^4 \times |D_e+F_e|_{J/\psi}$ & $9.83 \pm 0.29$ & $11.84 \pm 0.64$~\cite{Ferroli:2019nex,Ferroli:2020xnv} \\ %
$|D_e+F_e|_{\psi(2S)}$ & $3.31 \pm 0.55$ & $2.90 \pm 0.18$~\cite{Ferroli:2020mra} & $\sim 0.7$ \\ % 
%$10^4 \times |D_e+F_e|_{\psi(2S)}$ & $2.91 \pm 0.14$ & $2.90 \pm 0.18$~\cite{Ferroli:2020mra,Ferroli:2020xnv} \\ %  
$|D_e-F_e|_{J/\psi}$ & $3.7 \pm 1.3$ & $3.39 \pm 0.65$~\cite{Ferroli:2019nex} & $\sim 0.2$ \\ % 
$|D_e-F_e|_{\psi(2S)}$ & $0.81 \pm 0.57$ & $0.42 \pm 0.18$~\cite{Ferroli:2020mra} & $\sim 0.7$ \\ % 
\noalign{\smallskip}\hline
\end{tabular}
\end{table}
%%%
\begin{table} [H]
\vspace{-2mm}
\centering
\caption{Cross sections computed through Eq.~\eqref{eq.cs2}, using the values of Table~\ref{tab:comp-de-fe-real-j} (second column) according to the parametrization of Table~\ref{tab:A.par.gamma}.}
\label{tab:cs-all} 
\begin{tabular}{lrr} 
\hline\noalign{\smallskip}
$B \overline B$ & $\sigma_{B \overline B} (M_{J/\psi}^2) \ \rm [pb]$ & $\sigma_{B \overline B}(M_{\psi(2S)}^2) \ \rm [pb]$ \\
\noalign{\smallskip}\hline\noalign{\smallskip}
$p \overline p$ & $11.1 \pm 2.4$ & $1.38 \pm 0.47$ \\
$\Sigma^+ \overline \Sigma{}^-$ & $9.0 \pm 1.9$ & $1.23 \pm 0.42$ \\
$\Sigma^- \overline \Sigma{}^+$ & $0.91 \pm 0.62$ & $0.07 \pm 0.10$ \\
$\Xi^- \overline \Xi{}^+$ & $0.74 \pm 0.51$ & $0.07 \pm 0.09$ \\
\noalign{\smallskip}\hline
\end{tabular}
\end{table}
%%%
\begin{gather}
\label{eq.DepFe-val-real}
\begin{aligned}
|F_e|_{\jp} = (7.6 \pm 1.2) \times 10^{-4} \ {\rm GeV}\,, \nonumber \\
|F_e|_{\psii} = (2.06 \pm 0.56) \times 10^{-4} \ {\rm GeV}\,. \nonumber \\
\end{aligned}
\end{gather}
These values are in agreement with those obtained for the $\jp$ and for the $\psii$ in Refs.~\cite{Ferroli:2019nex} and~\cite{Ferroli:2020mra}, respectively, where, under the hypothesis of reality, they were found to be positive.\\
We can calculate, using Eq.~\eqref{eq.demfe} with $\rho=0$ and the values of $|D_e+F_e|$ and $|D_e|$ from Eqs.~\eqref{eq.De-val-ref} and~\eqref{eq.DepFe},
\begin{gather}
\label{eq.DemFe}
\begin{aligned}
|D_e-F_e|_{\jp} = (3.7 \pm 1.3) \times 10^{-4} \ \rm GeV\,, \\
|D_e-F_e|_{\psii} = (0.81 \pm 0.57) \times 10^{-4} \ \rm GeV\,. 
\end{aligned}
\end{gather}
%%%
\begin{figure}[H]
\includegraphics[width=0.99\columnwidth]{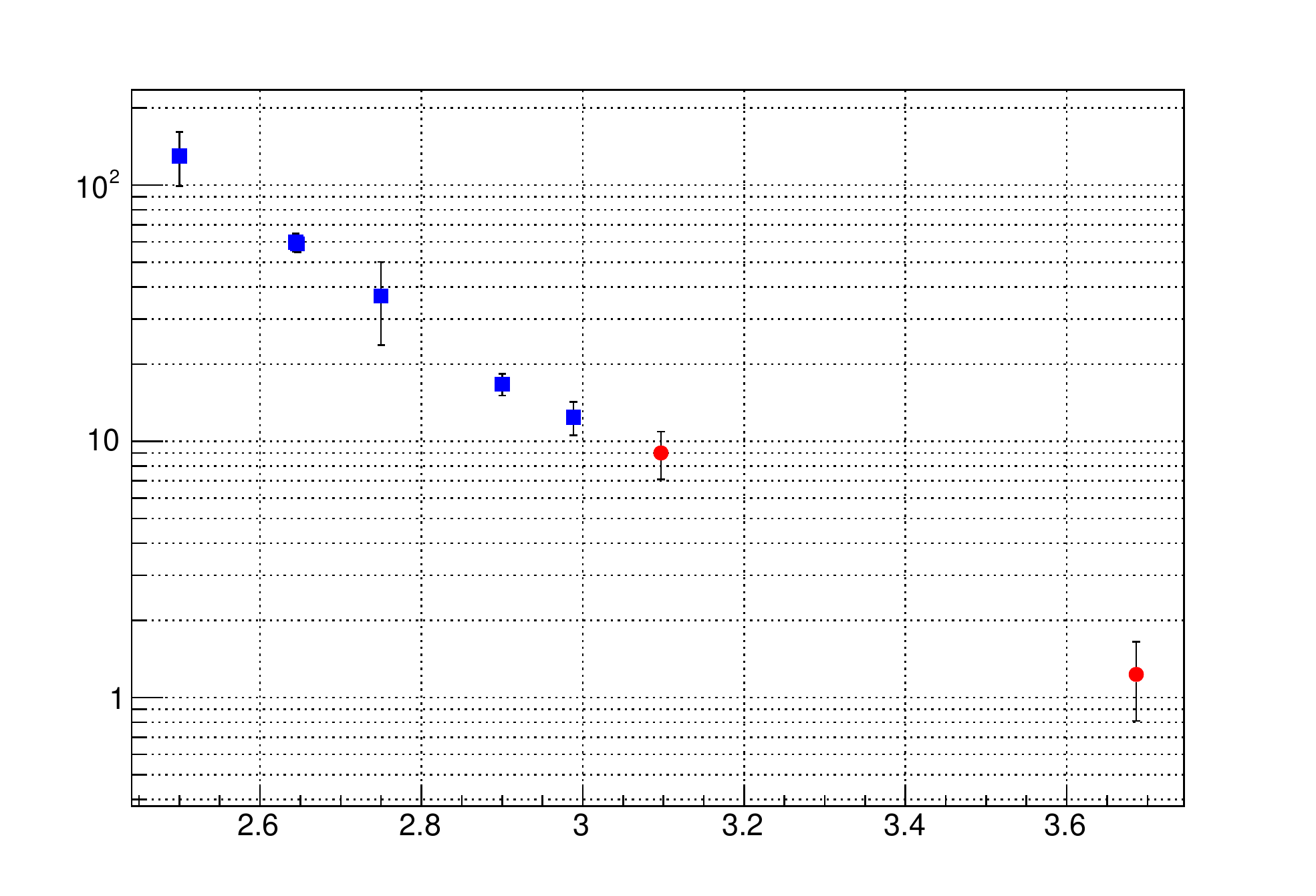}
\put(-220,157){$\sigma_{\Sigma^+ \overline \Sigma{}^-} \ \rm [pb]$}
\put(-67,-3){$\sqrt{q^2} \ \rm [GeV]$}\\
\includegraphics[width=0.99\columnwidth]{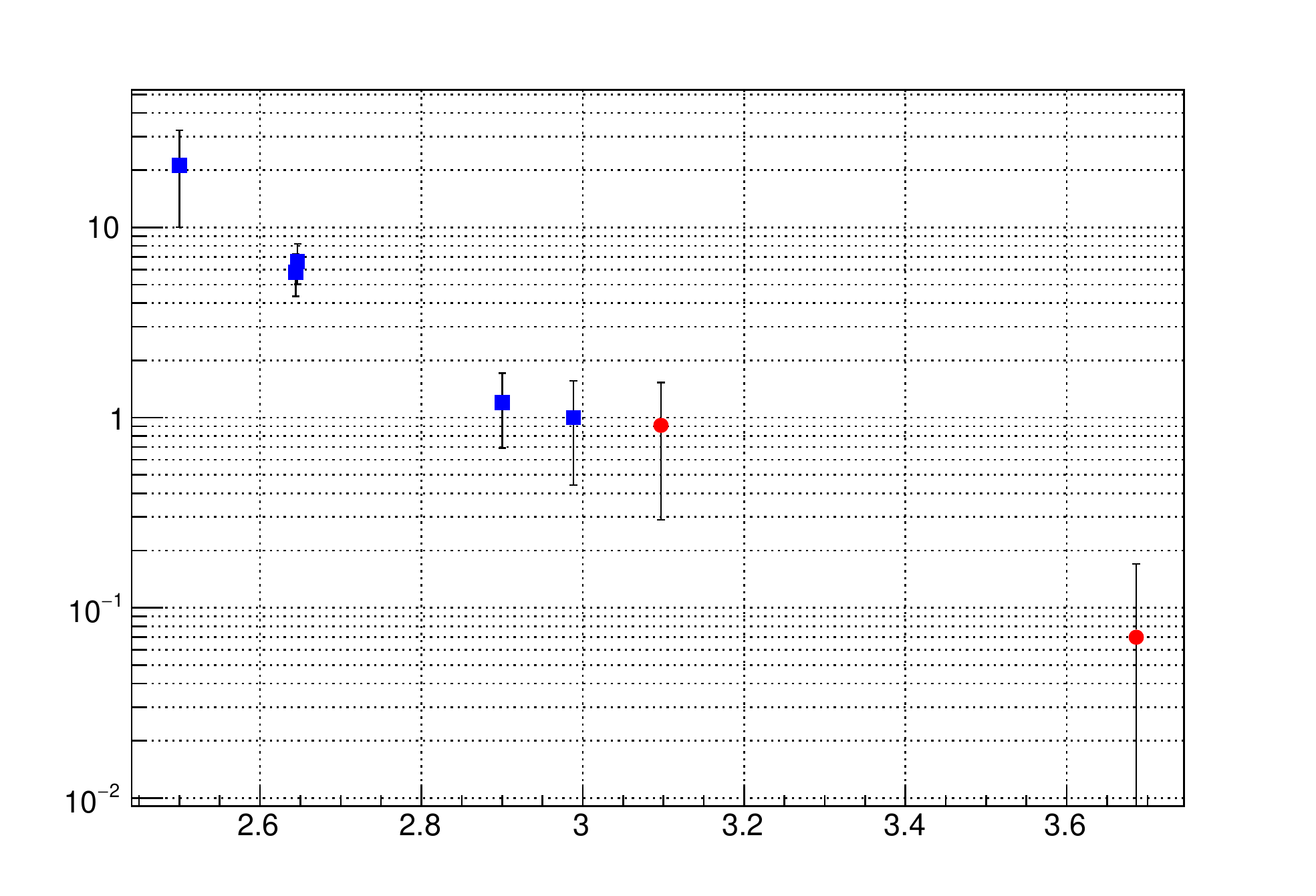}
\put(-220,157){$\sigma_{\Sigma^- \overline \Sigma{}^+} \ \rm [pb]$}
\put(-67,-3){$\sqrt{q^2} \ \rm [GeV]$}
\caption{\label{fig:sig-pm-cs} Plots of the $e^+e^- \to \Sigma^+ \overline \Sigma{}^-$ (upper panel) and $e^+e^- \to \Sigma^- \overline \Sigma{}^+$ (lower panel) cross sections in logarithmic $y$-scale. The blue squares are the data on the cross sections from BESIII~\cite{Ablikim:2020kqp}, while the red points indicate the values of the cross section at the $J/\psi$ and $\psi(2S)$ masses, reported in the second and third rows of Table~\ref{tab:cs-all}.}
\end{figure}
%%%
These values are reported, together with other quantities, in Table~\ref{tab:comp-de-fe-real-j}, where they are also compared with the values obtained in other works, all results are compatible within maximum 0.7 standard deviations.\\
Using the obtained values for $|D_e-F_e|$ of Eq.~\eqref{eq.DemFe} we can calculate the values of the cross sections of the other charged final state, $\sigma_{B \overline B}$, where $\BB \in \{ \Sigma^+ \overline \Sigma{}^-, \Sigma^- \overline \Sigma{}^+, \Xi^- \overline \Xi{}^+ \}$. The resulting values are reported in Table~\ref{tab:cs-all}.\\
In Figure~\ref{fig:sig-pm-cs} are shown the two obtained values of $\sigma_{\Sigma^+ \overline \Sigma{}^-}$ and $\sigma_{\Sigma^- \overline \Sigma{}^+}$, together with the available data on the cross section from the BESIII experiment~\cite{Ablikim:2020kqp}, where can be seen a good agreement with the power law decreasing trend as energy increases.

\section{Conclusions}
\label{sec:conclu}
We have considered the decays of a charmonium $\psi=J/\psi, \psi(2S)$ into a pair of spin-1/2 charged baryons $B$ to determine the complete parametrization of the EM amplitude using the BESIII data on the Born cross section $e^+ e^- \to p \overline p$.\\
The EM amplitudes of the decays $\psi \to \BB$ depend on two parameters, $D_e$ and $F_e$, and are proportional to the modulus $|D_e|$, for the neutral final states, and to $|D_e \pm F_e|$ for the charged ones.\\
We have previously obtained the value of $|D_e|$ at the $J/\psi$ and $\psii$ masses~\cite{Ferroli:2020xnv}, showing that the PDG value of the branching ratio for the so-considered purely EM decay $\psi \to \Lambda \overline \Sigma{}^0 + {\rm c.c.}$ is too large under this assumption.\\
In this work we have completed the determination of the moduli of all EM amplitudes for the decays $\psi \to \BB$, with the values of the parameter reported in Table~\ref{tab:comp-de-fe-real-j}.\\
The results agree within maximum 0.7 standard deviations with the predictions made in other works, where have been taken into account the full, strong, EM and mixed strong-EM amplitudes, of the decays $J/\psi \to \BB$~\cite{Ferroli:2019nex} and $\psii \to \BB$~\cite{Ferroli:2020mra}.\\
Using the results reported in Table~\ref{tab:comp-de-fe-real-j}, we have predicted the values of the cross sections $e^+ e^- \to \Sigma^+ \overline \Sigma{}^-$, $e^+ e^- \to \Sigma^- \overline \Sigma{}^+$ and $e^+ e^- \to \Xi^- \overline \Xi{}^+$, at the $J/\psi$ and $\psii$ masses, the values are reported in Table~\ref{tab:cs-all}. In particular, the values of the predicted cross sections for the $\Sigma^+ \overline \Sigma{}^-$ and $\Sigma^- \overline \Sigma{}^+$ final states are compared with those obtained by the BESIII collaboration at lower energies, as shown in Figure~\ref{fig:sig-pm-cs}, where the typical decreasing trend can be observed.
%
%
%\section*{Acknowledgement}
%%
%This work was supported in part by the STRONG-2020 project of the European Union's Horizon 2020 research and innovation programme under grant agreement No.~824093.
%
%

%\section*{Acknowledgement}
%%
%This work was supported in part by the STRONG-2020 project of the European Union's Horizon 2020 research and innovation programme under grant agreement No.~824093.

%\clearpage
% 


\begin{thebibliography}{}

%%%% NUOVI DA QUI

%\cite{Kopke:1988cs}
\bibitem{Kopke:1988cs}
L.~Kopke and N.~Wermes,
%``J/psi Decays,''
Phys. Rept. \textbf{174} (1989), 67
doi:10.1016/0370-1573(89)90074-4
%207 citations counted in INSPIRE as of 03 Oct 2020

%\cite{Claudson:1981fj}
\bibitem{Claudson:1981fj}
M.~Claudson, S.~L.~Glashow and M.~B.~Wise,
%``Isospin Violation in $J/\psi \to$ Baryon Anti-baryon,''
Phys. Rev. D \textbf{25} (1982), 1345
doi:10.1103/PhysRevD.25.1345
%62 citations counted in INSPIRE as of 03 Oct 2020

%\cite{Korner:1986vi}
\bibitem{Korner:1986vi}
J.~G.~Korner,
%``The $N - \Delta$ Transition Form-factor and Anomalous $\psi$ Decays Into Octet Decuplet Baryon - Anti-baryon Pairs,''
Z. Phys. C \textbf{33} (1987), 529
doi:10.1007/BF01548265
%16 citations counted in INSPIRE as of 03 Oct 2020

%\cite{Ferroli:2019nex}
\bibitem{Ferroli:2019nex}
R.~Baldini Ferroli, A.~Mangoni, S.~Pacetti and K.~Zhu,
%``Strong and electromagnetic amplitudes of the $J/\psi$ decays into baryons and their relative phase,''
Phys. Lett. B \textbf{799} (2019), 135041
doi:10.1016/j.physletb.2019.135041
[arXiv:1905.01069 [hep-ph]].
%4 citations counted in INSPIRE as of 02 Oct 2020

%\cite{Ferroli:2020mra}
\bibitem{Ferroli:2020mra}
R.~B.~Ferroli, A.~Mangoni, S.~Pacetti and K.~Zhu,
%``Amplitudes separation and strong-electromagnetic relative phase in the $\psi(2S)$ decays into baryons,''
Phys. Rev. D \textbf{103} (2021), 016005
doi:10.1103/PhysRevD.103.016005
%[arXiv:2005.11265 [hep-ph]].
%1 citations counted in INSPIRE as of 02 Oct 2020

%\cite{Ferroli:2020xnv}
\bibitem{Ferroli:2020xnv}
R.~B.~Ferroli, A.~Mangoni and S.~Pacetti,
%``The cross section of $e^+e^- \to \Lambda \overline \Sigma{}^0+{\rm c.c.}$ as a litmus test of isospin violation in the decays of vector charmonia into $\Lambda \overline \Sigma{}^0+{\rm c.c.}$,''
Eur. Phys. J. C \textbf{80} (2020) no.9, 903
doi:10.1140/epjc/s10052-020-08474-x
[arXiv:2007.12380 [hep-ph]].
%0 citations counted in INSPIRE as of 06 Oct 2020

% e^+e^- \to p \overline p BESIII
%\cite{Ablikim:2019njl}
\bibitem{Ablikim:2019njl}
M.~Ablikim \textit{et al.} [BESIII],
%``Study of the process $e^+ e^- \to p \bar p$ via initial state radiation at BESIII,''
Phys. Rev. D \textbf{99} (2019) no.9, 092002
doi:10.1103/PhysRevD.99.092002
[arXiv:1902.00665 [hep-ex]].
%13 citations counted in INSPIRE as of 02 Oct 2020

%\cite{Zyla:2020zbs}
\bibitem{Zyla:2020zbs}
P.~A.~Zyla \textit{et al.} [Particle Data Group],
%``Review of Particle Physics,''
PTEP \textbf{2020} (2020) no.8, 083C01
doi:10.1093/ptep/ptaa104
%175 citations counted in INSPIRE as of 03 Oct 2020

%\cite{Bianconi:2015vva}
\bibitem{Bianconi:2015vva}
A.~Bianconi and E.~Tomasi-Gustafsson,
%``Phenomenological analysis of near threshold periodic modulations of the proton timelike form factor,''
Phys. Rev. C \textbf{93} (2016) no.3, 035201
doi:10.1103/PhysRevC.93.035201
[arXiv:1510.06338 [nucl-th]].
%24 citations counted in INSPIRE as of 05 Oct 2020

%\cite{Shirkov:1997wi}
\bibitem{Shirkov:1997wi}
D.~V.~Shirkov and I.~L.~Solovtsov,
%``Analytic model for the QCD running coupling with universal alpha-s (0) value,''
Phys. Rev. Lett. \textbf{79} (1997), 1209-1212
doi:10.1103/PhysRevLett.79.1209
[arXiv:hep-ph/9704333 [hep-ph]].
%526 citations counted in INSPIRE as of 05 Oct 2020

%\cite{TomasiGustafsson:2001za}
\bibitem{TomasiGustafsson:2001za}
E.~Tomasi-Gustafsson and M.~P.~Rekalo,
%``Search for evidence of asymptotic regime of nucleon electromagnetic form-factors from a compared analysis in space- and time - like regions,''
Phys. Lett. B \textbf{504} (2001), 291-295
doi:10.1016/S0370-2693(01)00312-4
%45 citations counted in INSPIRE as of 05 Oct 2020

%\cite{Ablikim:2015vga}
\bibitem{Ablikim:2015vga}
M.~Ablikim \textit{et al.} [BESIII],
%``Measurement of the proton form factor by studying $e^{+} e^{-}\rightarrow p\bar{p}$,''
Phys. Rev. D \textbf{91} (2015) no.11, 112004
doi:10.1103/PhysRevD.91.112004
[arXiv:1504.02680 [hep-ex]].
%55 citations counted in INSPIRE as of 05 Oct 2020

%\cite{Lees:2013ebn}
\bibitem{Lees:2013ebn}
J.~P.~Lees \textit{et al.} [BaBar],
%``Study of $e^+e^- \to p \bar{p}$ via initial-state radiation at BABAR,''
Phys. Rev. D \textbf{87} (2013) no.9, 092005
doi:10.1103/PhysRevD.87.092005
[arXiv:1302.0055 [hep-ex]].
%113 citations counted in INSPIRE as of 05 Oct 2020

% e^+e^- \to p \overline p BABAR
%\cite{Aubert:2005cb}
\bibitem{Aubert:2005cb}
B.~Aubert \textit{et al.} [BABAR],
%``A Study of $e^{+} e^{-} \to p \bar{p}$ using initial state radiation with BABAR,''
Phys. Rev. D \textbf{73} (2006), 012005
doi:10.1103/PhysRevD.73.012005
[arXiv:hep-ex/0512023 [hep-ex]].
%274 citations counted in INSPIRE as of 02 Oct 2020

% e^+e^- \to \Sigma^+ \overline \Sigma{}^-
%\cite{Ablikim:2020kqp}
\bibitem{Ablikim:2020kqp}
M.~Ablikim \textit{et al.} [BESIII],
%``Measurements of Sigma+ and Sigma- Time-like Electromagnetic Form Factors for center-of-mass energies from 2.3864 to 3.0200 GeV,''
[arXiv:2009.01404 [hep-ex]].
%0 citations counted in INSPIRE as of 02 Oct 2020

%% e^+e^- \to \Xi^- \overline \Xi{}^+
%%\cite{Ablikim:2019kkp}
%\bibitem{Ablikim:2019kkp}
%M.~Ablikim \textit{et al.} [BESIII],
%%``Measurement of the cross section for $e^{+}e^{-}\rightarrow\Xi^{-}\bar\Xi^{+}$ and observation of an excited $\Xi$ baryon,''
%Phys. Rev. Lett. \textbf{124} (2020) no.3, 032002
%doi:10.1103/PhysRevLett.124.032002
%[arXiv:1910.04921 [hep-ex]].
%%0 citations counted in INSPIRE as of 02 Oct 2020



\end{thebibliography}
\end{document}